# Cytokine expression in malaria-infected non-human primate placentas

M. Barasa[1,2,*], Z.W. Ng'ang'a[3], G.A. Sowayi[1], J.M. Okoth[1], M.B.O. Barasa[4], F.B.M. Namulanda[5], E.A. Kagasi[2], M.M. Gicheru[2,6] and S.H. Ozwara[2]

[1]*Department of Medical Laboratory Sciences, School of Health Sciences, Masinde Muliro University of Science and Technology (MMUST), P. O. Box 190-50100, Kakamega, Kenya*
[2]*Department of Tropical and Infectious Diseases (TID), Institute of Primate Research (IPR) P. O. Box 24481-00502 Karen, Nairobi, Kenya*
[3]*Department of Medical Laboratory Sciences, Jomo Kenyatta University of Agriculture and Technology (JKUAT), Nairobi, Kenya P. O. Box 62000-00200, Nairobi. Kenya*
[4]*Center for Disaster Management and Humanitarian Assistance (CDMHA), Masinde Muliro University of Science and Technology (MMUST), P. O. Box 190-50100, Kakamega, Kenya*
[5]*College of Health Sciences, Moi Teaching and Referral Hospital (MTRH), Moi University, P. O. Box 4606-30100, Eldoret, Kenya*
[6]*Zoological Sciences Department, School of Pure and Applied Sciences (SPAS), Kenyatta University, P. O. Box 43844, Nairobi, Kenya*

___________________________________________________________________________________________

**Abstract**
Malaria parasites are known to mediate the induction of inflammatory immune responses at the maternal-foetal interface during placental malaria (PM) leading to adverse consequences like pre-term deliveries and abortions. Immunological events that take place within the malaria-infected placental micro-environment leading to retarded foetal growth and disruption of pregnancies are among the critical parameters that are still in need of further elucidation. The establishment of more animal models for studying placental malaria can provide novel ways of circumventing problems experienced during placental malaria research in humans such as inaccurate estimation of gestational ages. Using the newly established olive baboon (*Papio anubis*)-*Plasmodium knowlesi* (*P. knowlesi*) H strain model of placental malaria, experiments were carried out to determine placental cytokine profiles underlying the immunopathogenesis of placental malaria. Four pregnant olive baboons were infected with blood stage *P. knowlesi* H strain parasites on the one fiftieth day of gestation while four other uninfected pregnant olive baboons were maintained as uninfected controls. After nine days of infection, placentas were extracted from all the eight baboons through cesarean surgery and used for the processing of placental plasma and sera samples for cytokine sandwich enzyme linked immunosorbent assays (ELISA). Results indicated that the occurrence of placental malaria was associated with elevated concentrations of tumour necrosis factor alpha (TNF-α) and interleukin 12 (IL-12). Increased levels of IL-4, IL-6 and IL-10 and interferon gamma (IFN-γ) levels were detected in uninfected placentas. These findings match previous reports regarding immunity during PM thereby demonstrating the reliability of the olive baboon-*P. knowlesi* model for use in further studies.

___________________________________________________________________________________________

## Introduction

About 24 million pregnant women in Sub Saharan Africa are predisposed to PM and the incidences are higher in women in their first and second pregnancies (Phillips-Howard, 1999). Placental malaria causes adverse pregnancy outcomes like abortions, pre-term deliveries, still births, low birth weights (LBW), intrauterine growth retardation (IUGR) and reduction in gestational age. Furthermore, PM causes increased maternal and foetal mortality and morbidity (Menendez *et al*., 2000; Steketee *et al*., 2001).
Immunological mechanisms underlying the drastic pregnancy consequences during PM have been reported in human and animal model studies (Moore *et al*., 1999; Poovassery and Moore, 2006; Barasa *et al*., 2010a).

Different challenges are usually experienced during human PM studies such as inaccurate gestational age estimation, skepticism among study subjects, financial, ethical and moral obstacles (Steketee *et al*., 2001). Many questions concerning PM immunological mechanisms that lead to problematic pregnancies have not been clearly addressed due to these limitations (Steketee *et al*., 2001). One possible solution of avoiding these difficulties is the expansion of the number of feasible animal models, especially non-human primates (NHP) like olive baboons through extensive characterization experiments.

___________________________________________________________________________________________

***Corresponding Author:*** Mustafa Barasa, School of Health Sciences, Masinde Muliro University of Science and Technology, P. O. Box 190-50100, Kakamega, Kenya. *Tel:* +254 720442207. Email: *mustrech@yahoo.com*

The use of the recently established *P. knowlesi*-olive baboon (*Papio anubis*) PM model has increased the number of animal models available for placental malaria experiments (Barasa *et al*., 2010a; Ozwara *et al*., 2003).

This novel model of pregnancy/placental malaria makes use of the non - human primate and human malaria parasite *P. knowlesi* (Ng *et al*., 2008) whose entire genome has been sequenced (Ozwara *et al*., 2003) thus allowing the function of drug and vaccine candidate molecules to be screened using *P. knowlesi* genes (Ozwara *et al*., 2003).

The *P. knowlesi* parasite infects humans (Ng *et al*., 2008), is phylogenetically close to *P. vivax* and is highly virulent in the olive baboon system. Olive baboons are readily available and have immunopathophysiological and host-pathogen interactions that are similar to those in humans (Ozwara *et al*., 2003) making them ideal candidates for studying PM and or pregnancy associated malaria (PAM).

The above factors support the reliability of the components of the baboon-*P. knowlesi* model for pregnancy malaria studies. Therefore malaria investigations in NHP like olive baboons can be expected to provide insights and results that reflect what could happen in the human PAM and or PM conditions.

Placental malaria involves the preferential accumulation and sequestration of malaria parasites in the placental tissue where they induce pro-inflammatory immune responses that have been suggested to promote drastic PM incidences (Diouf *et al*., 2007). Despite extensive research in the area of PM and or PAM immunology, immunological mechanisms that render pregnant women susceptible to PM still require further clarification.

This study was undertaken due to this reason and also with the ultimate purpose of broadening the model options available in non-human primate pregnancy malaria experiments. Placental cytokine concentrations (TNF-α, IL-12, IL-4, IL-6, IL-10 and IFN-γ) of four pregnant olive baboons with PM were compared with those of four controls uninfected olive baboon placentas.

Findings communicated in this study provide new insights and results that are largely consistent with those documented from other animal model and human PM experiments, thereby increasing the relevance of the olive baboon-*P. knowlesi* PAM and or PM model in future investigations.

## Materials and Methods

### Acquisition and Maintenance of Experimental Olive Baboons (Papio anubis)

Eight adult female olive baboons (*Papio anubis*) originally captured from Kajiado District of Kenya with an average weight of 10 Kg were used in these experiments. All baboons were trapped and maintained in the quarantine facilities at the Institute of Primate Research, Karen, Kenya, for not less than three months. Examination of the baboons was done to confirm that they were free of hemoprotozoan, gastrointestinal parasites and Simian Immunodeficiency Virus (SIV) before inclusion into the study.

Microbiological examination of effusions, pus, ulcer material and skin specimens were done to confirm absence of microbial agents which cause infections in wounds and the skin. Before infection the baboons were maintained in the Institute of Primate research (IPR) at the Animal Resources Department's (ARD) baboon colony facilities.

All experiments were performed in accordance with the institutional guidelines of IPR as outlined by the IPR's Institutional Scientific and Ethical Review Committee (ISERC). Malaria infected baboons were transferred to the IPR biocontainment facility where each baboon was housed individually in a squeeze back cage, 0.6 x 0.6 x 0.68 meters high. They were maintained on a commercial non-human primate diet (Unga Millers Limited, Nairobi), supplemented with fruits, vegetables and additional ascorbic acid. Mineral salts and water were provided *ad libitum*.

The biocontainment facility was inspected for proper lighting, ventilation, drainage, temperature, foot bath sterilisation and sprayed with insecticide daily according to IPR standard operating procedures (SOP). For all invasive procedures, the baboons were anaesthetised with ketamine hydrochloride (10 mg/kg body weight).

### Experimental Design

The eight adult female olive baboons (*Papio anubis*) were maintained in the olive baboon colony facility of the IPR in the company of an adult male olive baboon for mating to occur. Ultrasonography was used to confirm pregnancy status and gestational levels of the olive baboons. Four of the olive baboons were infected with *P. knowlesi* blood stage, from parasites cultured overnight, on gestation day 150 (when the placentas were mature enough for experiments).

The other four olive baboons were maintained in uninfected state for use as control. On gestation day 159 all the eight olive baboons underwent cesarean surgery to facilitate the extraction of intact placentas for parasitaemia analysis and extraction of placental blood which was used for preparation of plasma (for TNF-α ELISA) and sera samples for IFN-γ, IL-4, IL-6, IL-10, IL-12 ELISA.

### In vitro Propagation of P. knowlesi and Infection of Olive Baboons

The *in vitro* propagation of malaria parasites was done using methods described previously (Ozwara *et al*.,

2003). Briefly, *in vitro* cultures were set-up using cryopreserved *P. knowlesi* parasites previously isolated from *P. knowlesi* infected olive baboons. Prior to culture, vials containing the parasite were removed from liquid nitrogen (-196°C) and quickly thawed at 37°C in a water bath andtransferred into 50 ml centrifuge tubes (Becton Dickinson). Subsequently, an equivalent volume of 3.5% NaCl (at room temperature (RT)) was added and mixed with the parasites before centrifugation at 800×g, (Sorvall RT 6000D) for 10 min at 24°C.

The supernatants were aspirated and 1/2 original volumes of 3.5% NaCl added before repeating the centrifugation. RPMI 1640 (Sigma, USA) with 10 % heat inactivated olive baboon serum (a volume equivalent to the first 3.5% NaCl) was added, mixed and centrifuged again as before. RPMI-10 (5x the original volume) was then added and mixed with the parasites for a final washing step by centrifugation as before. The parasite pellets were transferred into culture to a starting erythrocyte PCV of 2.5%.

The complete culture medium consisted of 2.5% baboon erythrocyte packed cell volume (PCV), 20% baboon serum, 15 µg/ml gentamycin solution and the rest RPMI 1640 (Invitrogen). All olive baboon sera for use had been heat inactivated previously at 56°C. One hundred to two hundred microlitres of culture were used for thin smear preparation (for determination of parasitaemia). Cultures were mixed gently and transferred into T25 culture flasks. The flasks were gassed, tightly capped and transferred to an incubator (37°C). Overnight cultured *P. knowlesi* parasites (Ozwara *et al.*, 2003) were used to initiate blood stage malaria infections in experimental baboons. $1.0 \times 10^6$ parasites/ml were resuspended in incomplete RPMI 1640 (Invitrogen). Each of the 8 sedated olive baboons received 1 ml of inoculum.

The parasites were injected through the femoral vein of the olive baboons' left legs via butterfly needles from 1 ml tuberculin syringes. Parasites were immediately flushed into the olive baboons' bodies using 5 ml saline solution for every baboon. Following infection, the olive baboons were carefully transported to the biocontainment facility and maintained in cages (Ozwara *et al.*, 2003).

*Cesarean Section and Parasitaemia Determination*
Caesarean deliveries were performed as detailed in the IPR SOP in order to obtain intact sterile placental tissue from olive baboons which could otherwise be consumed by them. Abdominal incisions were executed with minimum bleeding and light pressure manipulations resulted in the eventual expulsion of the placentas (Barasa *et al.*, 2010a).

The placentas were then placed in sterile dishes for extraction of placental blood. Each of the placentas was first rinsed once with 2% penicillin–streptomycin (P/S) in phosphate buffered Saline (PBS) (Diouf *et al.*, 2007).

Multiple incisions were made on the maternal side of the placenta and blood was squeezed out under sterile conditions. Thin placental blood smears were prepared and air dried for 5 minutes then fixed in 100% methanol. Fixed slides were stained for 10 min in 10% Giemsa solution and used for microscopic observation at x100 magnification. At least 2000 red blood cells (RBC) were counted in every parasitaemia determination session. This was done by counting erythrocytes from a quarter of each light microscope field and multiplying by four. Counting was done with the aid of manual lab counters. Parasitaemia was calculated as follows: Parasitaemia % = (Total counted parasites ÷ Total number of erythrocytes counted) x 100 (Ozwara *et al.*, 2003).

*Placental Plasma and Sera Preparation*
Blood samples for placental sera preparation were collected in sterile 50 ml tubes (Becton Dickinson). The blood samples were left at RT for 2 hours to allow for coagulation to take place. The blood samples were then subjected to overnight temperatures of 4°C. The blood was then spun at 850×g for 10 minutes at 24°C (Sorvall RT 6000D centrifuge).

Clear sera were sucked off and divided into 10 ml aliquots. Sera samples (for IFN-γ, IL-4, IL-6, IL-10, IL-12 ELISA) were stored at -20°C until use. For preparation of placental plasma samples for the TNF-α ELISA, a cytokine with a short half-life (Davison *et al.*, 2006), heparinised blood samples were centrifuged as above and the separated plasma samples were stored at -20°C (Ozwara *et al.*, 2003).

*Cytokine Sandwich ELISA*
These assays were performed essentially as described previously (Barasa *et al.*, 2010b; Barasa *et al.*, 2010c). Ninety-six well flat bottomed ELISA microtiter plates (Sigma-Aldrich, USA) were coated with 50 µl/well of cross reactive antihuman/baboon cytokine (IFN-γ, TNF-α, IL-4, IL-6, IL-10, IL-12) capture monoclonal antibody (Becton Dickinson) at a final concentration of 5 µg/ml.

After an overnight incubation at 4°C, excess coating buffer was flicked off and the wells blocked with 100 µl/well blocking buffer (3% BSA in PBS) followed by incubation at 37°C for 1 h. After washing the plates six times in ELISA washing buffer (0.05% tween in PBS (PBST (0.05%)), undiluted samples of either sera or plasma and recombinant cytokine standards were dispensed in duplicate (50 µl/well) and the plates incubated for at 37°C for 2h.

Standards were serially diluted by transferring 50 µl from well to well, commencing with a neat concentration of 500 pg/ml. The plates were then washed as before and detector mouse biotinylated antihuman/baboon cytokine (IFN-γ, TNF-α IL-4, IL-6,

IL-10, IL-12) monoclonal antibodies (Becton Dickinson) at dilutions of 1:2000 (50 μl/well). This was followed by incubation at 37°C for 1 h, and washed as before. Streptavidin Horse-Radish Peroxidase (SA-HRP; Sigma, USA) diluted 1:2000 was added 50 μl/well and incubated at 37°C for 1 h before a furthersix washes. Colour development was achieved by adding 50 μl/well of Tetramethylbenzidene (TMB) substrate (Sigma, USA) and optical densities were read using a Dynatech MRX ELISA reader at 630 nm filter setting after incubation at 37°C for 15 min.

*Statistical Analysis*

Mean values of cytokine concentrations in PM-infected olive baboon placentas were compared with the respective mean values of cytokine concentrations in malaria free olive baboon placentas using non parametric Mann-Whitney U analysis. Probability values of $P < 0.05$ were considered significant.

**Results**

*Placental Parasitaemia*

*P. knowlesi* parasites (Figure 1) were microscopically detected and quantified in the placentas of all the four infected olive baboons following cesarean surgery. Mean placental parasitaemia (14.4%) was over 20-fold higher than simultaneous mean peripheral parasitaemia (0.7%) on the cesarean section days. Microscopic examination of placental blood smears prepared from placentas extracted from uninfected control olive baboons confirmed the absence of parasites in these placentas (0.0% parasitaemia) hence their negative status with regard to placental malaria.

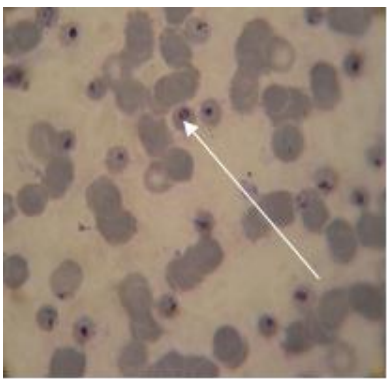

**Fig. 1.** A thin placental blood smear from a *P. knowlesi* infected baboon. Presence of malaria parasites was demonstrated in four baboon placentas in this study. The arrow points to a ring stage of *P. knowlesi* development.

*Placental Cytokine Responses*

The assayed cytokines were detectable in the placental sera (IFN-γ, IL-12, IL-4, IL-6 and IL-10) and plasma (TNF-α) samples extracted from malaria-infected and malaria-negative olive baboons (Figures 2 and 3).

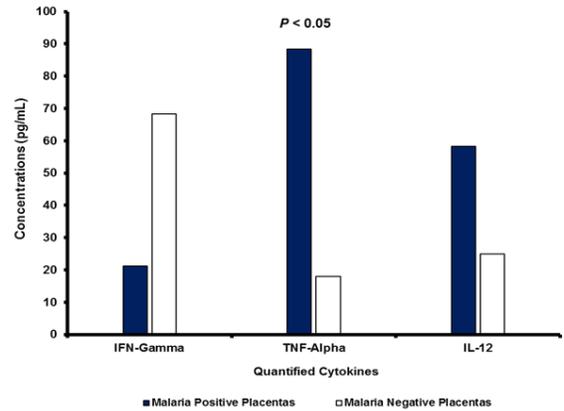

**Fig. 2.** Mean placental IFN-γ, TNF-α and IL-12 responses in *P. knowlesi* malaria positive and malaria negative placentas. With a mean concentration of 88.3 pg/mL, proinflammatory TNF-α was the most highly expressed induced innate cytokine in the malaria infected placentas. Concentrations of TNF-α in placentas positive for malaria were five times higher than in PM negative placentas ($P < 0.05$).

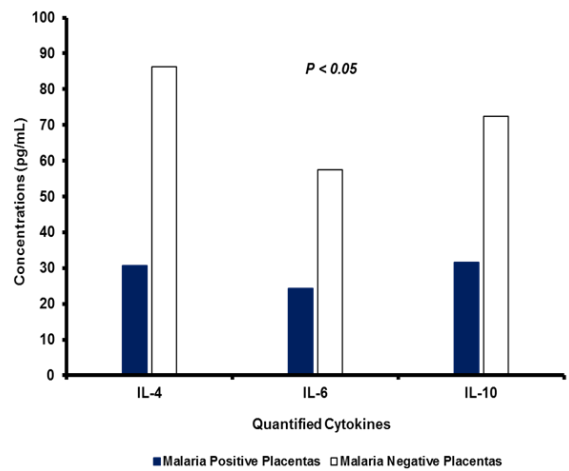

**Fig. 3.** Mean placental IL-4, IL-6 and IL-10 responses in *P. knowlesi* malaria positive and malaria negative placentas. Significantly higher concentrations of IL-4, IL-6, IL-10 and IFN-γ were detected in PM negative than in PM positive placental sera ($P < 0.05$). With a mean value of 86.3 pg/mL, IL-4 was the most highly detected anti-inflammatory cytokine in these experiments. The IL-4 levels in uninfected placentas were three times higher ($P < 0.05$) than in malaria infected baboon placentas.

The mean concentrations of IFN-γ cytokine were significantly lower ($P < 0.05$) in sera samples from *P. knowlesi* infected placentas (21.2 pg/mL) than concentrations in sera from uninfected placentas (68.3 pg/mL). TNF-α levels in plasma samples from malaria infected placentas (88.3 pg/mL) were significantly increased ($P < 0.05$) compared to levels in plasma samples from malaria negative placentas (17.9 pg/mL). TNF-α and IL-12 were detected at significantly ($P < 0.05$) lower levels in uninfected placentas in comparison with malaria infected placentas. Concentrations of IL-12 (58.3 pg/mL) were significantly increased ($P < 0.05$) in sera samples from malaria positive placentas compared to concentrations in sera samples from malaria negative placentas (24.9 pg/mL).

The mean concentrations of IL-4 cytokine were significantly lower ($P < 0.05$) in sera samples from *P. knowlesi* infected placentas (30.6 pg/mL) than concentrations in sera from uninfected placentas (86.3 pg/mL). IL6 levels in sera samples from malaria infected placentas (24.4 pg/mL) were significantly reduced ($P < 0.05$) compared to levels in sera samples from malaria negative placentas (57.5 pg/mL). Concentrations of IL-10 (31.6 pg/mL) were significantly reduced ($P < 0.05$) in sera samples from malaria positive placentas compared to concentrations in sera samples from malaria negative placentas (72.5 pg/mL).

**Discussion**

This study was designed to determine cytokine (IFN-γ, TNF-α, IL-12, IL-4, IL-6 and IL-10) immune responses that take place in the placental unit following PM infection in olive baboons (*Papio anubis*) infected with *P. knowlesi* H strain parasites. Parasite-mediated interference in the pregnancy-sustaining placental anti-inflammatory immunomodulation has previously been implicated as a mechanism that leads to PM-related complications (Diouf *et al*., 2007).

Our current findings suggest that there is a possible PM disease promotion role by TNF-α in the non-human primate host. Increased presence of TNF-α in malaria infected placentas was documented previously (Moormann *et al*., 1999). In that study which involved TNF-α mRNA quantification, increased expression in the placenta was also associated with IUGR. TNF-α was also previously confirmed as a trigger agent of abortions in the malaria infected primate host (Davison *et al*., 2006).

The previous reporting of occurrences of abortions in the *P. knowlesi* - olive baboon model of PM (Barasa *et al*., 2010a) and the current detection of elevated TNF-α expression in sera derived from PM infected placentas are findings that are in tandem with previous results (Davison *et al*., 2006). TNF-α is also a critical factor during murine PM that acts by inducing placental coagulopathy (Poovassery *et al*., 2009). In the present study, TNF-α levels in malaria infected placentas were 5 times higher than in malaria free placentas.

Reduced quantities of IFN-γ detected in the malaria infected placentas could have conferred increased susceptibility to PM. Significantly lower levels of the IFN-γ cytokine were reported to occur in the peripheral circulation of malaria-infected pregnant olive baboons compared to malaria-infected non-pregnant ones (Barasa *et al*., 2010c).

In human studies, higher concentrations of IFN-γ within the maternal foetal unit was previously implicated as critical protective factor against PM (Moore *et al*., 1999). In that study, intervillous blood mononuclear cells (IVBMC) from PM-negative multigravid women experienced elevated production of IFN-γ compared to IVBMC from PM-negative primigravid and secundigravid and PM-positive multigravid women following *in vitro* cytokine assays. IFN-γ was recently reported to occur at significantly higher concentrations in uninfected women with PM compared to PM positive women (Bayoumi *et al*., 2009). Our results confirm that the down-regulation of IFN-γ is strongly associated with presence of PM in the malaria infected host and that increased IFN-γ cytokine production in the placenta may be important for the onset of the protective responses against placental malaria. This report reveals that increased IL-12 values are associated with PM in olive baboons either as a risk factor or as a result of parasite-driven skewing. In accordance with our results, IL-12 producing monocytes were previously found to be significantly increased in *P. falciparum* infected placentas, compared to malaria free placentas (Diouf *et al*., 2007).

In this article, we surmise that a reduced IL-4 level is associated with the incidence of PM in the non-human primate (olive baboon) host. IL-4 was reported to be higher in concentration in uninfected placentas compared to infected placentas (Bayoumi *et al*., 2009). During pregnancy, the immune system may be biased towards antibody mediated immunity rather than towards cellular responses and this immunomodulation associated with successful pregnancies (Diouf *et al*., 2007). The significantly higher levels of IL-4 measured in uninfected baboon placentas is a finding that supports this phenomenon.

Malaria-free placentas had higher levels of IL-6 in comparison with infected placentas.

Decreased expression of IL-6 and transforming growth factor–b1 were found in malaria-infected placentas compared with uninfected placentas (Moormann *et al*., 1999). Significantly higher levels of IL-6 were measured in malaria - free placentas than in malaria infected placentas (Moormann *et al*., 1999). Control uninfected baboon placentas were discovered to have significantly elevated sera IL-10 levels. In another study (Bayoumi *et al*., 2009), elevated production of IL-10 was detected in malaria-free placentas compared to malaria-infected placentas. Our findings are in agreement with these previous results, indicating that increased levels of IL-10 are may be needed for protective responses against PM. The data presented here provide new insights into the cytokine concentration changes that take place in the non-human primate model of PM.

Uninfected olive baboon placentas expressed significantly higher levels of IL-4, IL-6 and IL-10 compared to malaria infected placentas, indicating that higher levels observed in this panel of cytokines (Diouf *et al*., 2007; Poovassery *et al*., 2009) were maintained in uninfected placentas.

This phenotypic polarization in the malaria free placentas became altered towards the TNF-α and IL-12-dominated cytokine panel following placental malaria. The presence of *P. knowlesi* parasites in infected baboon placentas is therefore revealed in these experiments as a factor that leads to the onset of an inflammatory immune response in the placental unit. This could potentially be detrimental to foetal development.

Further information from these experiments has shown that IFN-γ has a crucial protective role against PM since reduced levels of this cytokine were detected in the malaria infected placentas. Cytokine responses in the placentas were significantly altered following the onset of PM leading to a shift in immunity towards IL-12 and the induced innate TNF-α domination. This communication is an addition to the available knowledge base required in the development of immunological control measures. Immunological parameters associated with placental malaria in this NHP model were found to correspond with those described in the human situation indicating that the *P. knowlesi*-olive baboon model of pregnancy/placental malaria is suitable for further malaria immunology studies.

**Acknowledgments**
These experiments were funded by the research capability strengthening World Health Organisation (WHO) grant (Grant Number: A 50075) for malaria research in Africa under the Multilateral Initiative on Malaria/Special Programme for Research and Training in Tropical Diseases (WHO-MIM/TDR). We are grateful to all members of the ARD and the Reproductive Health and Biology Department at the IPR for providing and maintaining the olive baboons and performing cesarean surgeries during the study.

_________________________________________